\documentclass{article}
\usepackage[a4paper, margin=1.2in]{geometry}
\usepackage{authblk}
\usepackage{amsmath}
\usepackage{graphicx}
\usepackage[hyphens]{url}
\usepackage{booktabs}

\title{From NEA and NIA to NESAS and SCAS: Demystifying the 5G Security Ecosystem}
\author[1]{Angelos Michalas}
\author[2,3]{Constantinos Patsakis}
\author[2]{Dimitrios D. Vergados}
\author[4]{Dimitrios J Vergados}

\affil[1]{Department of Electrical and Computer Engineering, University of Western Macedonia, Kozani, Greece}
\affil[2]{Department of Informatics, University of Piraeus, Greece}
\affil[3]{Information Management Systems Institute of Athena Research Centre, Greece}
\affil[4]{Department of Informatics, University of Western Macedonia, Kastoria, Greece}

\begin{document}
\date{}
\maketitle

\begin{abstract}
Despite the numerous pompous statements regarding 5G, it is indisputable that 5G creates a radical shift in telecommunications. The main reason is that 5G is an enabler of numerous applications we have long envisioned and either simulated or implemented in test environments, partially or on a smaller scale. 5G will soon unlock the potential of smart cities, industry 4.0, and IoT, to name a few. However, a crucial question is how much we can trust this technology. Since this technology will soon become the core infrastructure for all of the above, it is critical to understand the fundamental security mechanisms that comprise this technology and the guarantees they provide to assess the potential risks we are exposed to. This work follows a non-technical yet bottom-up approach to introduce the reader to the core security mechanisms and establish a baseline for the security of 5G, to demystify the principal notions and processes. Based on the above, we streamline future directions and highlight possible threats.
\end{abstract}

\section{Introduction}
Consumers see resilient networks as a vital help in coping with everyday life. With more and more activities being carried out online, and greater numbers of hours spent connected to mobile broadband and the connection to the internet have become an indispensable part of daily life, remote working, education/e-learning and wellness, online healthcare consultations and telecare is expected to become more popular than physical visits to the doctor – as critical as access to food and electricity.

Over the past few years, there have been extensive discussions about the fifth generation of cellular networks, 5G. While people consider it a simple upgrade from the existing 4G networks, the truth is that it is far more than that. In principle, the speeds exceed wired networks at a latency below five milliseconds, practically allowing mobile devices to provide real-time feedback. As a result, 5G is an enabler for many technologies involving IoT, smart cities, ubiquitous computing and industry 4.0. Therefore, 5G delivers higher performance, efficiency, and quality, ultimately leading to new user experiences and industry uptake. The above justifies that 5G is not just another hype but a step towards realising a plethora of services and products that would be seamlessly delivered worldwide.

High-speed data transmission and low latency are inherent features of 5G networks, reducing the time between actions and responses, making devices, machines and sensors more responsive, and enabling real-time two-way communication and remote control. Ultimately, private networks can propagate over 5G, allowing machines, tools, parts, and personnel on production lines to be fully synchronised, improving performance and enabling features and coverage characteristics that facilitate mass customisation.

5G is intended to complement or completely replace the existing 4G LTE mobile network. Each mobile network generation is defined by several factors, including the technology used, the delay (i.e. the time between sending and receiving a signal), and the overall speed of data transmission over the network to the end device. Two of the most interesting developments driven by the introduction of 5G are the implementation of Open-RAN (O-RAN) and network slicing capabilities.

On the one hand, Open Radio Access Network, commonly referred to as Open RAN, is a concept that focuses on interoperability and standardisation of radio access network elements. This includes a unified interconnection standard for white-box hardware and open-source software from different vendors. The Open RAN architecture integrates a modular base station software stack on off-the-shelf hardware, allowing other suppliers' baseband and radio unit components to work together seamlessly. It emphasises the streamlined performance objectives of 5G RAN through the common attributes of efficiency, intelligence, and versatility. Open RAN can benefit 5G applications such as autonomous systems, support network slicing, and enable secure and efficient over-the-air firmware upgrades when deployed at the network edge. This is because the edge provides the low-latency and high-reliability requirements that these applications need. 

On the other hand, network slicing refers to dividing a single cellular network into multiple virtual networks. This allows multiple virtual networks to exist on a common shared physical infrastructure. These virtual networks can be customised to meet the specific needs of applications, services, devices, customers, or operators. This way, each slice can have an independent set of logical network functions, allowing it to better fulfil the requirements of a specific use case without compromising the usability of other slices and allowing the reuse and sharing of common resources. Thus, network slicing maximises 5G network flexibility, optimising both infrastructure utilisation and resource allocation \cite{7926920}.

5G will affect not only the competitiveness of our communications sector but the competitiveness of potentially every sector and every so-called “vertical” industry, making it the cornerstone of digital transformation. Telecommunication networks are evolving rapidly across a broad technological environment, including virtualisation, IoT and Industry 4.0. This is met by an equally broad yet deteriorating cybersecurity environment. Advances in technology, together with the broader development of networks beyond 5G RAN, are expected to significantly impact security, such as software-defined networking (SDN), network function virtualisation (NFV), and edge computing.

The most exciting implementation of 5G seems to be the upgrade of IoT devices that offer enhanced network capabilities. 5G will connect more devices with faster speeds and very low latency. Integrating 5G into his IoT devices could significantly improve the user experience while enabling virtual and augmented reality applications, regardless of the device or service used. Moreover, the digitisation of all the traditional main economic sectors, such as agriculture, manufacturing and services, is expected to accelerate with the 5G realisation. The implementation of the 5G network in the European Union will format the future of European society and economy and will have a significant impact on the lives of EU citizens. Billions of products and systems are connected in every sector of the economy, including energy, transportation, industrial systems, banking and health. Processes such as elections are also increasingly based on digital infrastructure and 5G networks.

The number of connected IoT devices is expected to reach 14.4 billion by the end of this year \cite{stateofiot}. Notably, the number of Massive IoT connections increased by a factor of 3 in 2019, reaching close to 100 million. The Massive IoT technologies NB-IoT and Cat-M1 continue to be rolled out around the world Industries are taking steps to incorporate connectivity and cellular into their standards, as many industrial enterprises are defining 5G as their primary connectivity platform for both IT and OT systems to reach new levels of productivity, security, and safety. Manufacturers see 5G as a new platform for their operational technology (OT), eventually through dedicated resources to ensure critical manufacturing processes are guaranteed the connectivity resources they require. Enterprises provided vital input to 3GPP in developing the IMT-2020 (5G) standards, resulting in cellular networks designed for their needs. Industry bodies now combine membership from both manufacturing and ICT companies, for example, with 5GAA in automotive and 5G-ACIA in industry. The Critical Communications Association (TCCA) pulls together stakeholders in the public safety arena. In the newly emerging field of air traffic management for beyond-line-of-sight (BVLOS) drones, bodies such as NASA and FAA in the US, EASA in the EU, and the Global Unmanned Aircraft Systems Traffic Management Association (GUTMA) work on standards, and 3GPP follows with work items to align. For live broadcast production (e.g. news gathering, sports coverage), the European Broadcasting Union (EBU) has a working group for 5G in Content Production (5GCP), while 3GPP studies the requirements of audio and video production.

According to GSMA, the adoption of 5G in Europe is significantly lagging compared to the USA, China, Japan, and South Korea. Even if the majority of European countries (34 out of 50) have already deployed 5G, and more than half of operators (92 out of 173) have launched 5G networks, only 2.5\% of the connections are 5G when the corresponding connections are 14.2\% in the USA and almost double that (28\%) in the area of China, Japan, and South Korea. While there are many reasons for the lack of adoption by 2025, it is estimated that there will be 1.2 billion connections globally.

The 5G Action Plan aimed at reaching the 5G objectives of a coordinated launch (quasi-simultaneous) in all Member States by the end of 2020 and a rapid build-up in urban areas, socio-economic driver zones and along all main transport paths (roads and railways) by 2025, security and privacy are most important in the 5G full deployment will contribute towards realistic AI-based security management. This work aims to reveal the security challenges linked to the greater access of third-party suppliers to networks and interlinkages between 5G networks and third-party systems contributing directly to upcoming policy work on accountability, certification, and liability. The next decade will see significant technological developments toward the Next-Generation Internet (NGI) in a global competition context. It will be fuelled by Artificial Intelligence and the cloudification/softwarisation of network components. 5G standards are expected to evolve regularly to reflect these technological advances, up to a point where the planning of the next generation, 6G, is likely to become a concrete prospect. Therefore the goal is to impact the trustworthiness of this NGI race.

On October 9 2019, the EU Member States, with the support of the European Commission and the European Cybersecurity Agency, published a high-level report on a coordinated risk assessment of 5G networks. This completed an important step in the European Commission’s Recommendation on Common EU Concepts for 5G Network Security. The report is based on national risk assessments submitted to the Commission by all Member States earlier this year. It identifies the top cyber threats and actors, the most sensitive assets, the top vulnerabilities, and a set of strategic risks. With 5G revenues worldwide estimated at €225 billion by 2025, 5G will be a key factor for Europe to compete in the global market, and increased cybersecurity will ensure that the EU retains technological sovereignty.

Given the evolving nature of 5G technology and its environment, we report a number of security challenges that are likely to be created or exacerbated by the advent of 5G networks. 5G network technologies and standards will also introduce specific security improvements compared to previous network generations. Still, due to the new characteristics of network architecture and the wide range of services and applications that may rely heavily on 5G networks in the future, some new important security issues may also arise. More specifically:

\begin{enumerate}
\item The technological changes introduced by 5G will increase overall attacks. The number of gateways on the surface and possible attackers:
    \begin{itemize}
        \item Increased functionality and less centralised architecture compared to previous generations of mobile networks. This means that some core network functions can be integrated into other parts of networks, increasing the sensitivity of corresponding devices, like base stations.
        \item  Increasing the software share in 5G devices leads to increased risk in the software development and update processes.
    \end{itemize}
\item These new technological features make the reliance of mobile network operators on third-party suppliers even greater, which will, in turn, increase the number of attack paths that malevolent parties could exploit. In this context, a supplier’s individual risk profile will be important, especially when a supplier has a presence within networks or areas.
\item Heavy reliance on a single supplier increases the risk and exposure and consequences of a failure of this supplier.
\item If 5G networks become a vital part of the supply chain of IT and its applications, confidentiality and privacy requirements will be impacted. Furthermore, the integrity of the networks will become a matter of intercommunal as well as international security concerns from an EU perspective.
\end{enumerate}

As a result, 5G security is a hot and timely topic which requires special attention from the whole ecosystem. While undoubtedly there is a lot of work in this direction, both theoretical and practical, due to its massive scope, many aspects of 5G security remain unclear for practitioners who simply want to use this technology and understand the security guarantees one can have from adopting 5G. To fill this gap, this work tries to provide a generic and not very technical overview of the cybersecurity 5G ecosystem, following a bottom-up approach, contrary to following a stakeholder-based \cite{SURACI2021107604} which focuses more on the attacks and risks but not the underlying security mechanisms. To this end, we present each layer's security features, measures, and relevant legal frameworks that try to make 5G secure and resilient.

The rest of this article is structured as follows. In the next section, we present the related work, focusing on describing the 5G ecosystem and stakeholders, the vertical ecosystems, and the main security drawbacks of 5G’s predecessors. Then, in Section 3, we discuss the regulatory frameworks, focusing more on the NIS directive and the GDPR. Section 4 discusses the main network security primitives, privacy measures, vendor-wise security processes and certification, and real-world attacks. Finally, we summarise the article and discuss the road ahead for 5G security.

\section{Related work}
\subsection{5G ecosystem and stakeholders}
As the latency and capacity of the 5G network improved the online experience dramatically, new actors and stakeholders emerged who are responsible for developing the 5G technology, providing 5G connectivity, and using the capabilities of the 5G network to improve and even revolutionise entire industries significantly. As the new stakeholders and use cases that emerge are often very critical to the operation of society, and also the new players of the 5G ecosystem increasingly depend on the 5G network for their day-to-day operation, we will present an overview of the 5G ecosystem and the stakeholders who participate in it.

\begin{figure}[th]
    \centering
    \includegraphics[width=\textwidth]{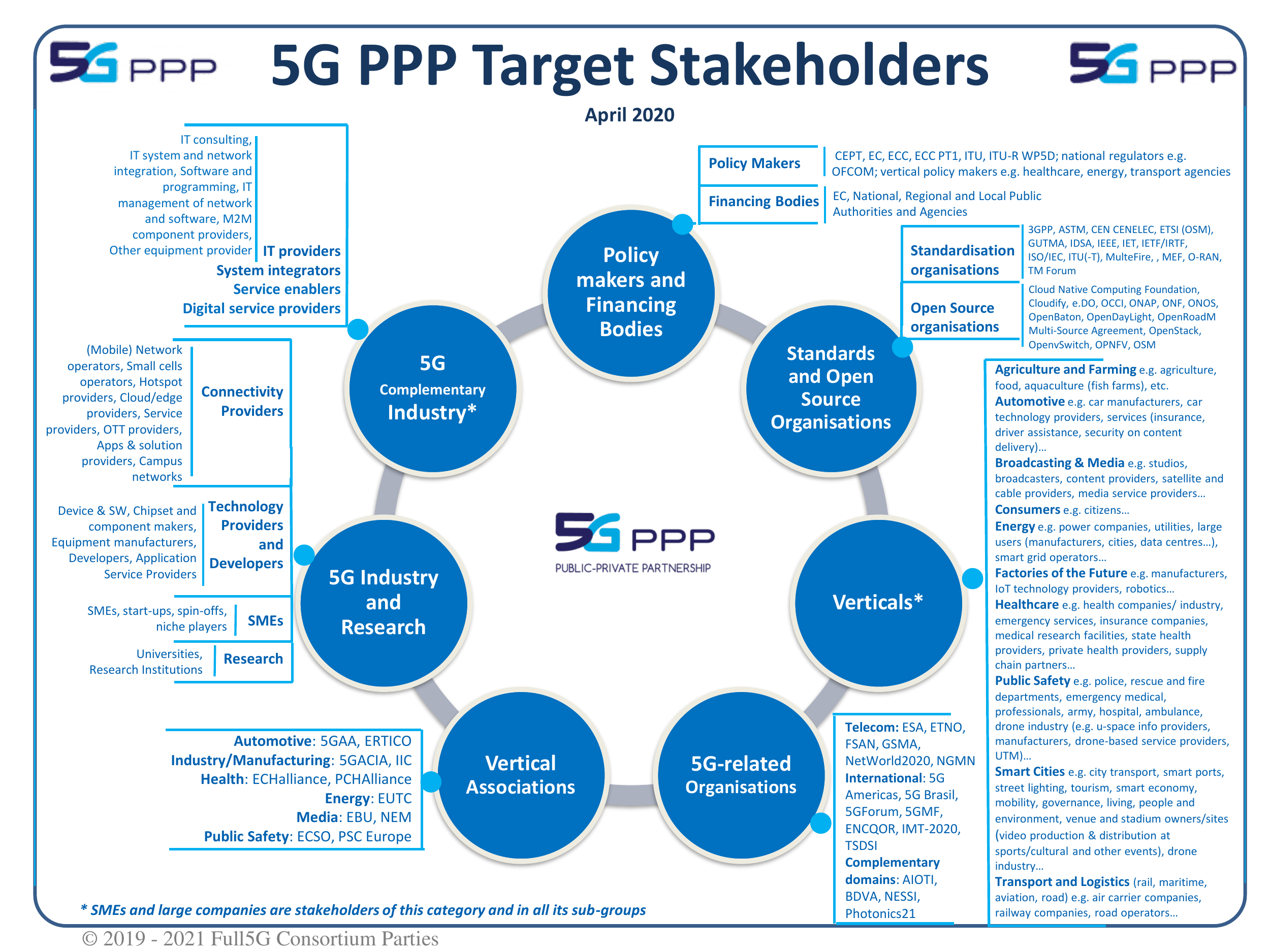}
    \caption{5G Stakeholders according to 5G PPP \cite{magen20205g}.}
    \label{fig:5g_stakeholders}
\end{figure}

A comprehensive overview of the stakeholders that participate in the 5G ecosystem has been conducted by 5G-PPP \cite{magen20205g}, see Figure \ref{fig:5g_stakeholders}. The main clusters of these actors that are identified are the following:
\begin{itemize}
    \item Policy makers and Financing Bodies, who decide the overall policy for 5G adoption and also provide financing to the other stakeholders.
    \item Standards and Open Source Organisations, and stakeholders who define the various standards for the different protocols and interfaces used in the 5G ecosystem and also develop the open software that is utilised in these solutions.
    \item Verticals represent industrial sectors that benefit from and contribute to the 5G ecosystem
    \item 5G – related organisations, which are either telecommunication providers or international associations 	
    \item Vertical Associations are specific associations focused on developing a specific industrial sector.
    \item 5G Industry and Research encompass the various connectivity providers, technology providers, and developers, as well as SMEs and research institutes
    \item 5G Complementary Industry, includes stakeholders such as IT providers, System Integrators, Service enablers and Digital service providers.
\end{itemize}
However, the above can be complemented by non-traditional stakeholders due to the verticals. A typical example can be considered car manufacturers who are continuously integrating more connectivity features in their vehicles, utility providers who want to implement smart grids and connectivity via 5G can provide many solutions for them, local/national governments for implementing smart cities, or even small appliances vendors for IoT, and prominent industry players and industrial equipment manufacturers who aim to integrate 5G connectivity for industry 4.0.

\begin{figure}[th]
    \centering
    \includegraphics[width=.8\textwidth]{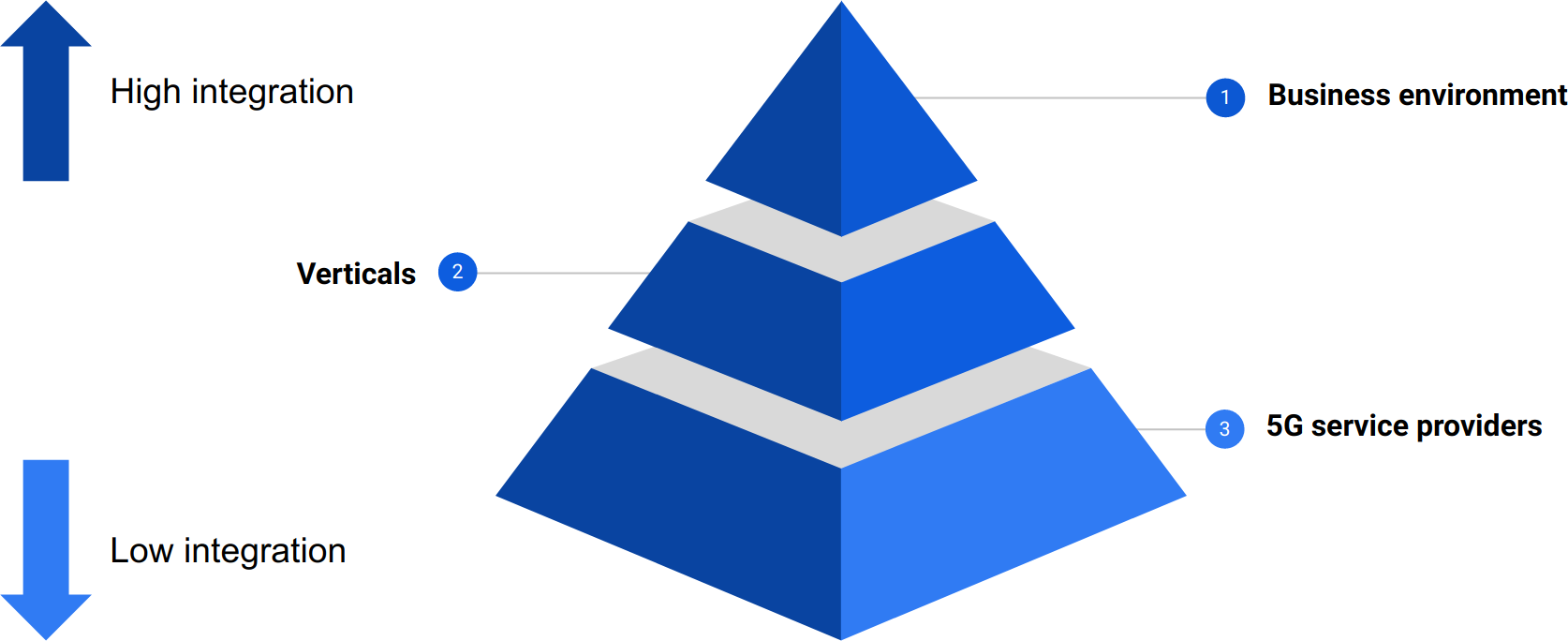}
    \caption{Layers for analysing 5G ecosystems according to the integration level.}
    \label{fig:layers}
\end{figure}

Another categorisation of the above-mentioned stakeholders into three layers (see Figure \ref{fig:layers}), depending on their level of integration in the 5G ecosystem. The layers are the 5G service providers, the verticals, and the Business environment. The stakeholders who are central in the standardisation, development, and provisioning of the 5G infrastructure comprise the lower layer of the ecosystem, with the various verticals that rely on the 5G system comprising the second layer and the general businesses that will benefit from 5G adoption comprising the upper layer. All the mentioned layers are highly dependent on the security mechanisms of the 5G network, but their role is in the development and use of these mechanisms.

\begin{figure}[th]
    \centering
    \includegraphics[width=\textwidth]{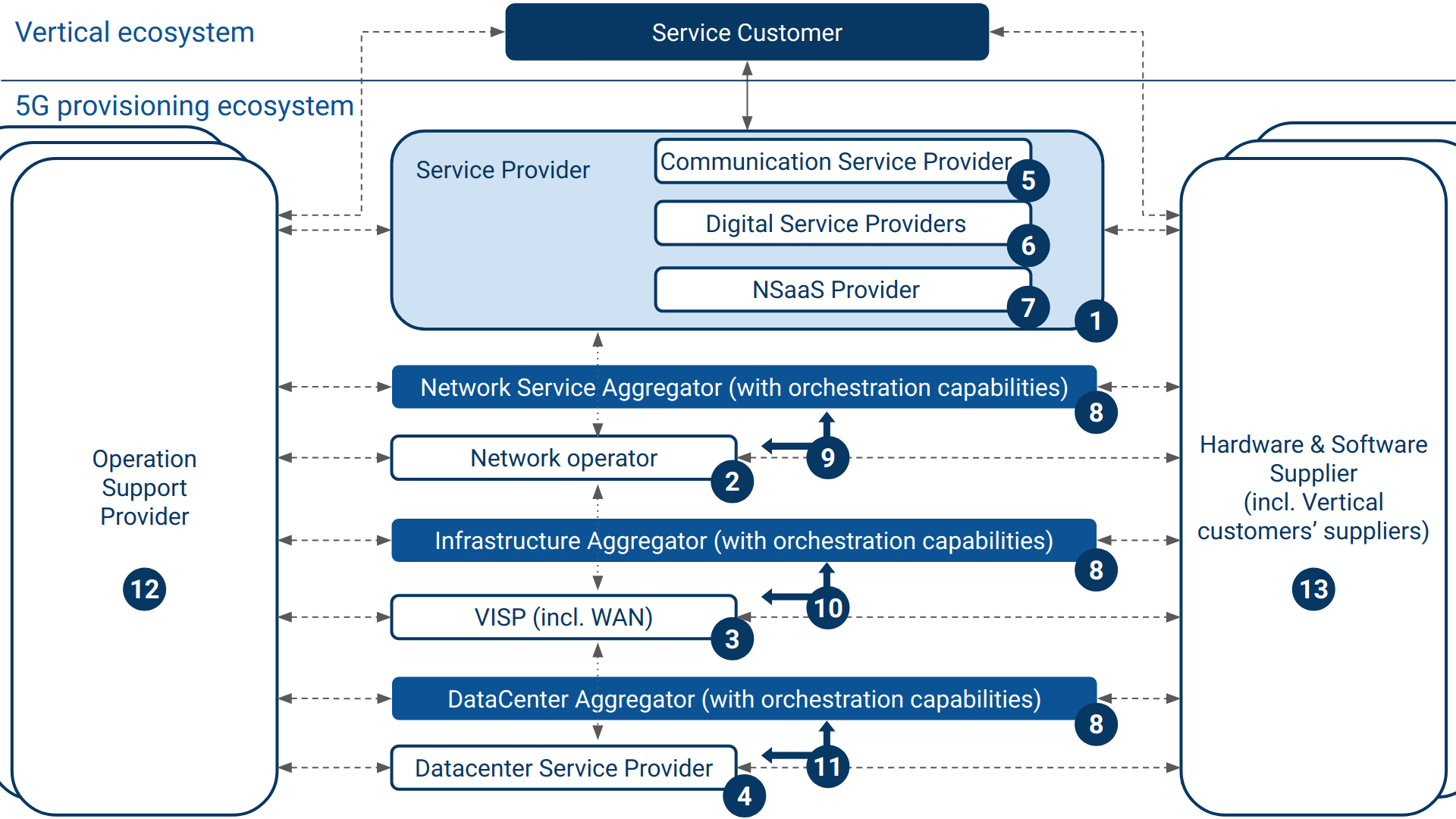}
    \caption{Roles in 5G provisioning systems (5GPPP) \cite{redana20195g}.}
    \label{fig:roles}
\end{figure}

The stakeholders are also differentiated according to their role in the 5G ecosystem. From the point of view of the stakeholders that are providing the 5G service to the end-users, i.e., those with a provisioning role. The provisioning roles in 5G are summarised in Figure \ref{fig:roles}; adapted from \cite{redana20195g}. The principal role in the 5G system is the role of the service provider (1), who obtains and orchestrates resources for the Network Operators (2), Virtualisation Infrastructure Service Providers (VISP) (3) and the DataCenter Service Providers (DCSP) (4) The Service Provider includes the roles of the Communication Service Provider (5), the Digital Service Provider (6), and the Network Slice as a Service Provider (7). Additionally, the corresponding aggregator roles are identified (8), as well as the roles of the Operation Support Provider (12) and the Hardware and Software Providers (13). All these stakeholders are integral components in the provisioning of a secure 5G system, as discussed further in the following sections.

\subsection{The 5G vertical ecosystems}
\begin{figure}[th]
\includegraphics[width=\textwidth]{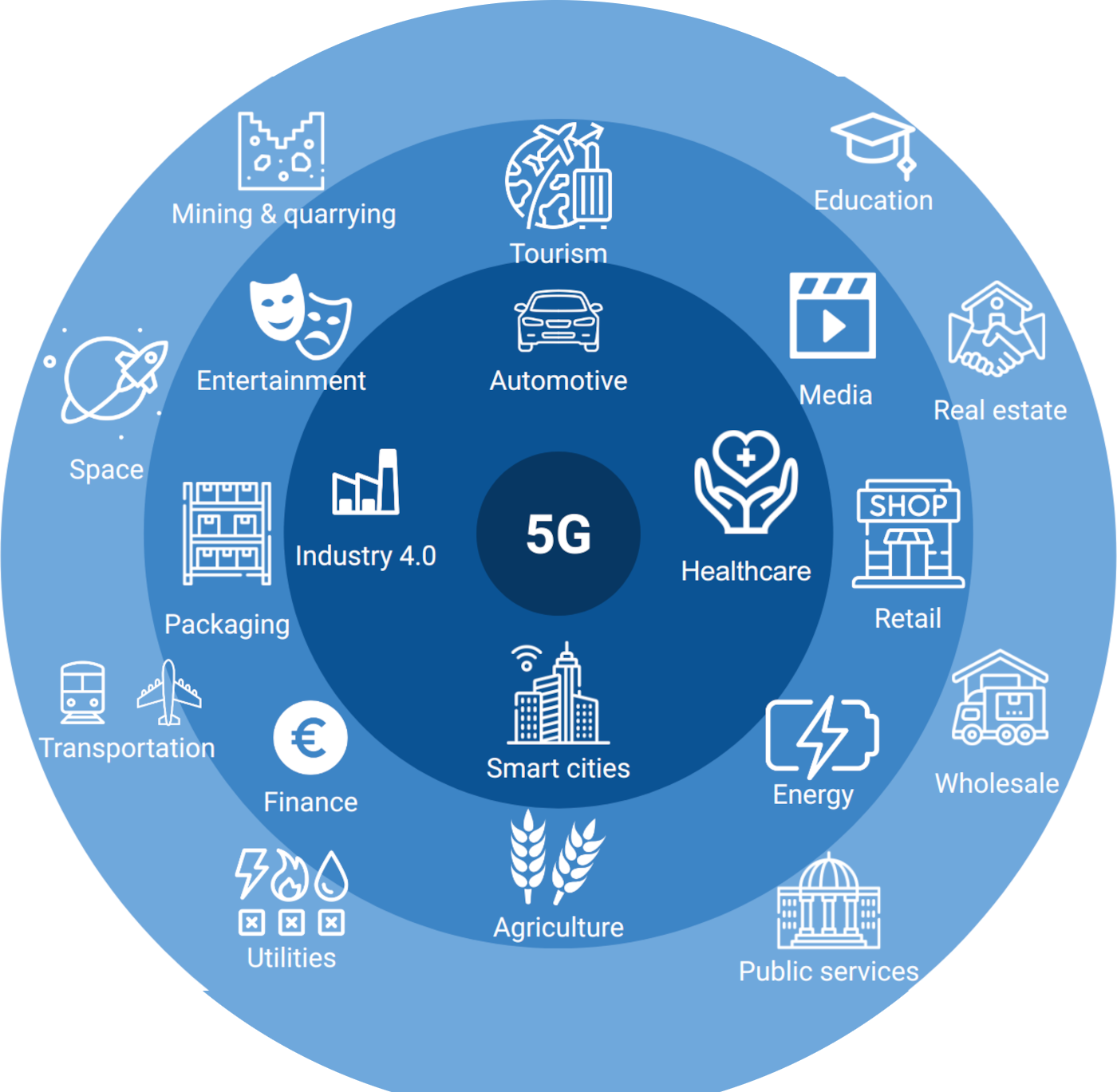}
    \caption{A non-exhaustive mapping of the 5G vertical ecosystem.}
    \label{fig:verticals}
\end{figure}

A new concept that is important in the security standardisation of the 5G network is the concept of the verticals, i.e. the industry sectors that will benefit and contribute to the 5G ecosystem. For these sectors, 5G can enable new business opportunities. The specific security requirements of each vertical may be different among the different verticals. Still, all of them are tightly dependent on the 5G ecosystem, and thus the reliability and security of the network are of extreme importance to them. These vectors greatly vary and continuously expand as new applications emerge. Some of these verticals are very explored, e.g. automotive, transportation, media, Smart Cities, healthcare, factories of the future, energy, public safety, tourism, agriculture, forestry and fishing, storage, wholesale, and retail sales, to name a few. Figure \ref{fig:verticals} illustrates some of these verticals.

\section{Regulatory frameworks}
In what follows, we present the main two regulatory frameworks in the EU for 5G, namely the GDPR and NIS directive.
\subsection{The General Data Protection Regulation}
The General Data Protection Regulation (GDPR) \cite{gdpr} is a comprehensive data protection law that applies to all individuals within the European Union (EU). It was put into effect on May 25, 2018, and replaces the Data Protection Directive \cite{dpd}, which had been in place for more than 20 years. The GDPR is designed to strengthen and harmonise data protection laws across the EU while also addressing privacy concerns that have arisen in the digital age. The GDPR applies to the processing of personal data by organisations operating within the EU, as well as by organisations outside the EU that offer goods or services to individuals within the EU. It sets out specific rules for the collection, use, and storage of personal data and gives individuals certain rights in relation to their personal data, including the right to access, correct, erase, and restrict the processing of their data.

The GDPR also introduces new obligations for organisations to report certain data breaches to relevant authorities and affected individuals and imposes significant fines for non-compliance. Overall, the GDPR is intended to give individuals greater control over their personal data and to enhance the protection of their privacy rights. The GDPR includes several principles that aim to give individuals more control over their personal data, including the right to object to profiling, the right to data portability, and the obligation for organisations to conduct data protection impact assessments. One of the most widely debated rights introduced by the GDPR is the Right to be Forgotten, which is specified in Article 17 of the regulation. This right allows individuals to request the deletion of their personal data when there is no longer a valid reason for it to be processed. This right has been a subject of controversy \cite{politou2018forgetting}, as it raises questions about the balance between an individual's right to privacy and the public's right to access information.

The GDPR is a technology-agnostic law, meaning that it does not recommend specific technical frameworks or privacy-preserving methods for compliance. Instead, it outlines functional requirements at a high level of abstraction, leaving it up to organisations to decide how best to implement them using the technologies available. Theoretically, this approach allows the GDPR to remain relevant and applicable regardless of current trends and state-of-the-art technologies in the field of computer science. By not binding the provisions of the law to specific technologies, legislators were able to ensure that the GDPR could adapt to future technological developments. Nevertheless, this may lead to potential issues \cite{politou2018backups,kutylowski2020gdpr} when trying to implement it as there might be practical limitations which may couple with the lack of baseline technologies and monitoring mechanisms and prevent the provision of the desired functionality and impact. 

As discussed, 5G has many verticals. While 5G may have many well-established security mechanisms (see the following sections), the GDPR may significantly impact the development and deployment of 5G in relation to data protection. The GDPR mandates organisations to implement appropriate technical and procedural measures to protect personal data from unauthorised access, use, storage, handling, and disclosure. Given the diversity of 5G stakeholders and the complexity of this infrastructure and systems, errors leading to vulnerabilities and lack of establishing proper measures among them are two well-known risks.

Moreover, the GDPR may impact 5G in relation to data privacy. The GDPR gives individuals certain rights concerning their personal data, including the right to access, correct, erase, and restrict the processing of their data. Organisations will need to ensure that they can continuously respect and uphold these rights in 5G services. Consider the case of the automotive vertical. Thus, we assume that a car manufacturer wants to develop a connected car that uses 5G to provide a range of services to the driver and passengers through, e.g. the infotainment system. These services may simultaneously involve real-time traffic updates, personalised recommendations for routes and destinations, and in-car entertainment based on their interactions. Clearly, to provide these services, the car manufacturer will need to process personal data, such as the driver's location, preferences, and search history, along with affective features extracted from the passenger's reactions in real-time and for possibly millions of other users. The amount of collected data and its type are subject to the protections and rights outlined in the GDPR as, among others, they contain sensitive personal data such as biometrics and the individual's location data. Similarly, a manufacturing company may want to implement Industry 4.0 technologies, from sensors, robots, and advanced analytics, to, e.g. improve its efficiency, productivity, and product quality. Using 5G to realise these technologies implies the transmission and processing of data which may involve the processing of personal data, such as health data of the employees, their whereabouts, their financial status, bank account, their home addresses etc., but also the same for customers, and suppliers.

Let us consider for the time being that the service provider for the two verticals above, the car manufacturer and the smart factory, have the informed and direct consent of the data subject to collect, process, and store the corresponding information. Given that a big part of 5G is virtualised and ported to the cloud, we have to assess the user exposure to passive adversaries in the 5G ecosystem. In this context, one has to assess per installation the possible direct exposure of user data to an honest but curious VISP or DCSP. Moreover, given the capacities of encrypted traffic analysis \cite{taylor2017robust,conti2018dark,casino2019hedge,rezaei2019deep,aceto2019mobile,liu2019fs,aceto2021distiller} would it be possible for a Network Slice as a Service Provider or some of the aggregators (discussed in the related work) to infer user actions and use of resources from the data they process? Have all the possible user tags been properly anonymised to avoid user exposure, and are all the data processed and stored in the user-designated jurisdiction?

Clearly, the above questions cannot be answered straightforwardly and are subject to configurations performed by the corresponding providers requiring thorough and in-depth assessment to determine their compliance with the GDPR. The interested reader may also refer to \cite{teatini2019privacy}, and \cite{rizou2020gdpr}.

\subsection{The NIS directive and the upcoming NIS 2.0}
\begin{figure}[th]
    \centering
    \includegraphics[width=\textwidth]{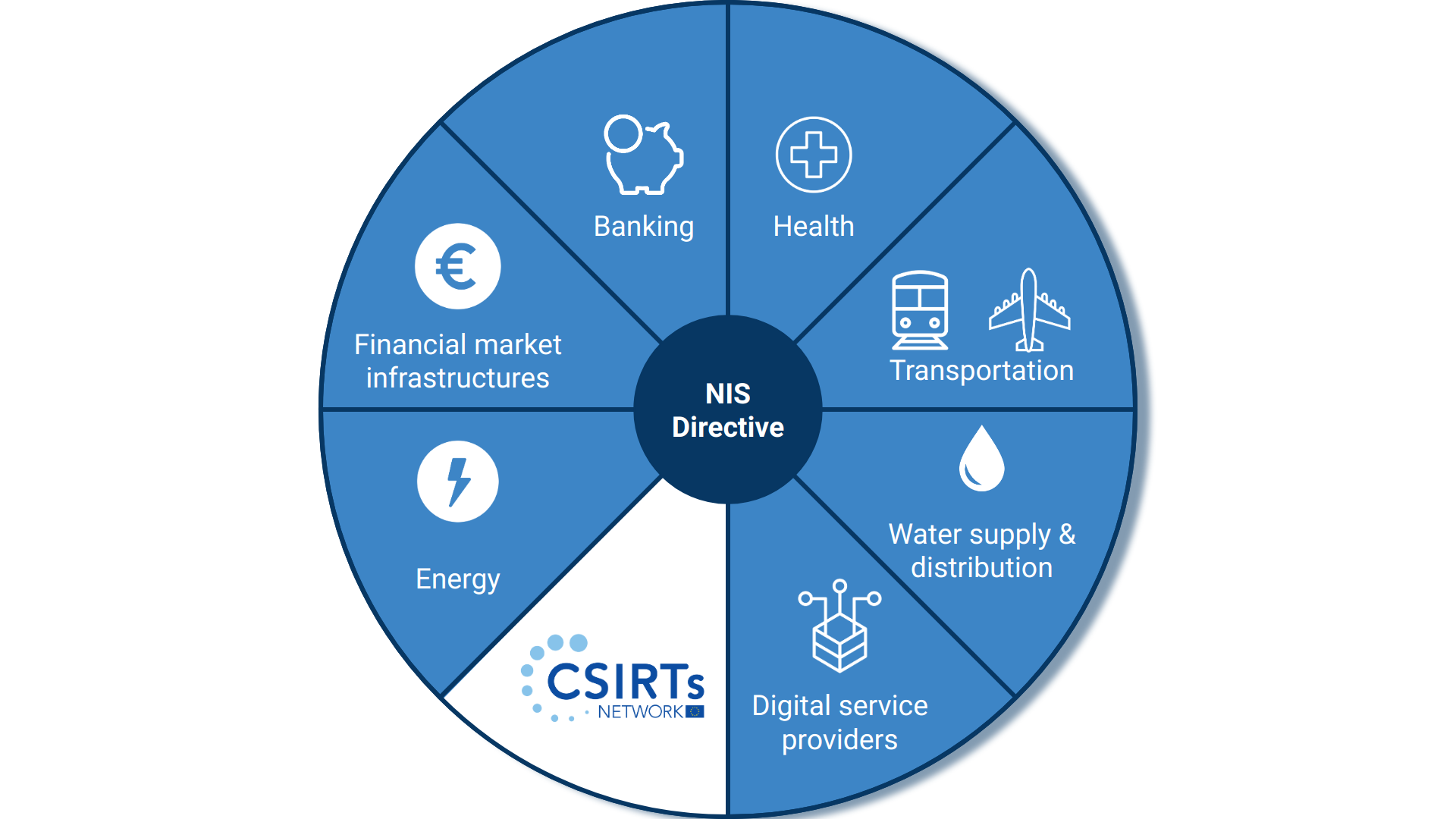}
    \caption{NIS directive sectors.}
    \label{fig:nis_sectors}
\end{figure}

To specify the steps required to achieve a high level of security for networks and information systems, primarily involving essential service operators (OES) and digital service providers, the Directive (EU) 2016/1148 \cite{european2016directive}, also known as the NIS Directive, was introduced to EU Law. To improve the overall level of cybersecurity in the EU and support the smooth operation of the internal market, the NIS Directive offers legal measures, and it is founded on three major pillars:
\begin{itemize}
    \item The NIS Directive mandates that the Member States adopt a national strategy on the security of network and information systems to ensure that they are highly prepared. Member States must also establish national Computer Security Incident Response Teams (CSIRTs), which act as a single point of contact, an effective national NIS authority, and risk and incident management teams. To ensure cross-border cooperation between Member State authorities and the relevant authorities in the other Member States, as well as with the NIS Cooperation Group, each CSIRT must serve as a liaison.
    \item The CSIRTs Network, which encourages quick and efficient operational cooperation between national CSIRTs, and the NIS Cooperation Group, which supports and facilitates strategic collaboration and the exchange of information among the Member States, are both established by the NIS Directive.
    \item The NIS Directive ensures cybersecurity measures are implemented across seven mission-critical sectors, including energy, transport, banking, financial market infrastructures, drinking water, healthcare, and digital infrastructures, see Figure \ref{fig:nis_sectors}. NIS covers all sectors considered crucial for the European economy and society and is heavily reliant on ICT.
\end{itemize}
Notably, this is the first piece of EU-wide cybersecurity legislation.

Public and private organisations designated by the Member States as providers of essential services (OES) in these industries are required to assess their cybersecurity risks and implement reasonable and proportionate security controls. Serious incidents must be reported to the appropriate authorities. Additionally, digital service providers (DSPs) who offer important digital services like search engines, cloud computing, and online marketplaces must abide by the directive's security and notification requirements. Meanwhile, the latter are governed by a so-called "light-touch" regulatory regime, which includes, among other things, placing them under the control of a single Member State for the entirety of the EU and exempting them from ex-ante supervisory measures.

The scope of application of OESs and DSPs was not sufficiently defined in the directive, leaving the Member States with too much latitude in implementing the requirements and identifying the entities involved. As a result, some clarity issues were discovered. Additionally, it was determined that the sectors' "coverage" was too narrow because the directive could no longer accurately represent all industries that provide crucial services to the economy and society.

In addition to the aforementioned, the European Commission has acknowledged the need for a regulatory framework to address the significant increase in cyber security risks brought on by recent digitisation and greater interconnection. For instance, the rise in the use of cloud services following the widespread adoption of remote working due to the Covid-19 pandemic led to a surge in cyber-attacks \cite{cisa,lallie2021cyber,europol2021internet,deloite}. Moreover, the introduction of 5G technologies, and IoT devices, among other factors, imply new measures that were not initially foreseen.

While NIS significantly increased the cybersecurity capabilities of the Member States, its implementation was challenging as cyberspace continuously evolves, and new threats emerge. Thus, the Commission will update and upgrade the NIS Directive to strengthen the security requirements, address the new security challenges for supply chains, and define how reporting will be made, by whom and when, including harmonised sanctions across the EU. As a result, NIS 2.0 will oblige more entities and sectors to take measures and increase the level of cybersecurity in the EU. 

According to the first directions of NIS 2.0\footnote{\url{ https://www.europarl.europa.eu/thinktank/en/document/EPRS_BRI(2021)689333}}, the main changes are the following: 
\begin{itemize}
\item Based on the importance of new industries (such as wastewater management, food, and space, among others) to the economy and society, the scope of the directive is broadened to include all medium-sized and large businesses operating in these industries. At the same time, flexibility in identifying smaller entities with a high-risk profile is guaranteed to the Member States;
\item To support the coordinated management of cybersecurity on significant incidents and crises at the EU level, the creation of a European Cybercrises Liaison Organization Network (EU-CyCLONe) is proposed;
\item The disclosure of newly discovered vulnerabilities across the Union is more closely coordinated;
\item The new proposal eliminates the distinction between OES and DSP, instead classifying entities as either essential or important;
\item The coverage of the directive is expanded to cover new sectors (e.g. wastewater management, food, space and so on) based on their criticality for the economy and society, including, for this purpose, all medium and large companies of these sectors. At the same time, Member States are guaranteed flexibility in identifying smaller entities with a high-risk profile;
\item The establishment of a European Cyber Crises Liaison Organisation Network (EU-CyCLONe) is proposed to support the coordinated management of cybersecurity on large-scale incidents and crises at EU level
\item Greater coordination is established in the disclosure of new vulnerabilities discovered throughout the Union;
\item A list of administrative sanctions (similar to those of the GDPR) is established, including fines for violating cybersecurity risk reporting and management obligations;
\item The proposal strengthens security requirements for businesses by enforcing a risk management approach and providing a minimum list of basic security elements that must be applied. In addition, it introduces more precise provisions on the process of reporting incidents, the content of the reports and the timing (within 24 hours of the discovery of the incident);
\item The proposal introduces stricter supervisory measures for national authorities, stricter enforcement requirements, and aims to harmonise sanctioning regimes across the Member States;
\item At the European level, the proposal strengthens cybersecurity for key information and communication technologies. In cooperation with the Commission and ENISA, the Member States will have to carry out coordinated risk assessments of critical supply chains, building on the effective approach taken in the context of the Commission's Recommendation on the cybersecurity of 5G networks.
\end{itemize}

With the revision of the NIS directive; therefore, the European Commission proposes a re-elaborated version of the level of cyber security throughout the Union to increase the resilience of the various sectors involved, both in public and private spheres.

While under the old NIS directive, member states were responsible for determining which entities would meet the criteria to qualify as operators of essential services, the new NIS 2.0 directive introduces a size-cap rule. This means that all medium-sized and large entities operating within the sectors or providing services covered by the directive will fall within its scope.

While the agreement between the European Parliament and the Council maintains this general rule, the provisionally agreed text includes additional provisions to ensure proportionality, a higher level of risk management and clear-cut criticality criteria for determining the entities covered.

The text also clarifies that the directive will not apply to entities carrying out activities in areas such as defence or national security, public security, law enforcement and the judiciary. Parliaments and central banks are also excluded from the scope.

As public administrations are also often targets of cyberattacks, NIS 2.0 will apply to public administration entities at the central and regional levels. In addition, member states may decide that it applies to such entities at the local level too.

A stronger governance architecture for cybersecurity in Europe will be a central feature of NIS 2.0. ENISA (European Network Information Security Agency), the European Commission and the 27 EU member states will work more closely together in drawing up risk assessment and mitigation strategies covering the security of supply chains in Europe. The key requirement businesses want concerning the rollout and implementation of NIS 2.0 is certain. Companies want to make decisions knowing exactly what the rules of engagement are in the context of NIS 2.0.

As the NIS 2.0 proposal now stands, EU member states have a lot of discretion in deciding how to enact NIS 2.0. Countries can invoke non-technical criteria in carrying out an assessment of a prospective security risk to the European supply chain. While the current geopolitical situation has gravely reshaped the cybersecurity landscape, we argue that the place or origin of a supplier should not be a criterion that should be permitted to be included as a risk to the European supply chain unless there are specific constraints, e.g. the recent sanctions after the Russian invasion of Ukraine.

ENISA, the European Commission, and the 27 member states of the EU should draw up harmonised technical rules that will clearly lay down a zero-trust approach in strengthening the security of European supply chains. This will also play an important role in avoiding fragmentation in the effective operation of the internal market in Europe, as noted by EU representatives \cite{killeen_2022}, especially under future NIS 2.0 arrangements. The building of a zero-trust approach for an effective rollout of NIS 2.0 across Europe should be based on the principles of openness, transparency and non-discrimination.

Such a strategy will give small, medium and large-scale European businesses the certainty they need in complying with the spirit and legal obligations of NIS 2.0. The bottom line is that supply chain risk should be based on hard evidence and approved and certified international standards. Cybersecurity is, of course, a global challenge. It would be in the interest of the European Union to push for international cybersecurity standards in the area of supply chain risk. Securing such an agreement between the EU and other relevant international organisations will provide higher levels of cybersecurity and avoid unnecessary bureaucratic burdens.

The NIS 2.0 Directive significantly expands the current scope of application, which is of major importance. Growing interconnectedness, rapid digitisation and ubiquitous connectivity mean more sectors are becoming systemically important to defend from cyber risk than before. 

The new rules have added seven more industries in both the public and private sectors to its scope.

\begin{itemize}
\item Providers of public electronic communications networks or services: In 2020 alone, the EU member states reported a total of 170 telecom security incidents resulting in 841 million lost user hours and gathering support for the NIS 2.0 directive. Till now, European telecom operators notified “significant” security incidents to their national authorities, who would compile a report for the EU Cybersecurity Agency at the end of each year. 
\item Waste and water management: The nature of public water systems and their limited operational technology usage make them attractive to attackers as they have the potential to harm the environment, economies and citizens. The inclusion of waste disposal and water management fall under the EU’s new directives seems to be a wise precautionary measure.
\item Manufacturers of certain critical products: The new directive sets out to cover the healthcare sector more broadly than before by including medical device and pharmaceutical manufacturers in its scope, though there are concerns that it might overlap existing medical device regulations.
\item Food: The new cybersecurity obligations will now cover all medium to large-sized enterprises in the food distribution sector as a ‘critical’ sector.
\item Digital services such as social networking platforms and data centers: Social media platforms will have to make incident reports, and if the company is based outside the EU, then it has to keep a representative in the EU to address grievances. Incident report rules also apply to the domain name, service providers, and data banks.
\item Space: The European Union Agency for Space Programme will also fall under the ambit of the directives, as will its member agencies to a certain degree. The NIS 2.0 Directives recommend the EUASP have some influence on how satellite data management policies under the NIS 2.0 directives will operate.  
\item Postal and courier services: Europeans’ increasing dependence on postal services during the pandemic has seen the EU cover its members’ postal and courier entities under its cybersecurity framework. Postal company servers contain sensitive information such as names, addresses, contact details, etc., and may pose an attractive target to threat actors.
\end{itemize}
Standard rules for public administration websites: So far, individual member governments have been maintaining their own cybersecurity infrastructure, with the 2016 NIS recommendations serving as a standard but not a mandate, but now all member states’ administration websites and servers need to abide by the same rules. 

\section{Assessing the security of 5G networks}
\begin{figure}[th]
    \centering
    \includegraphics[width=\textwidth]{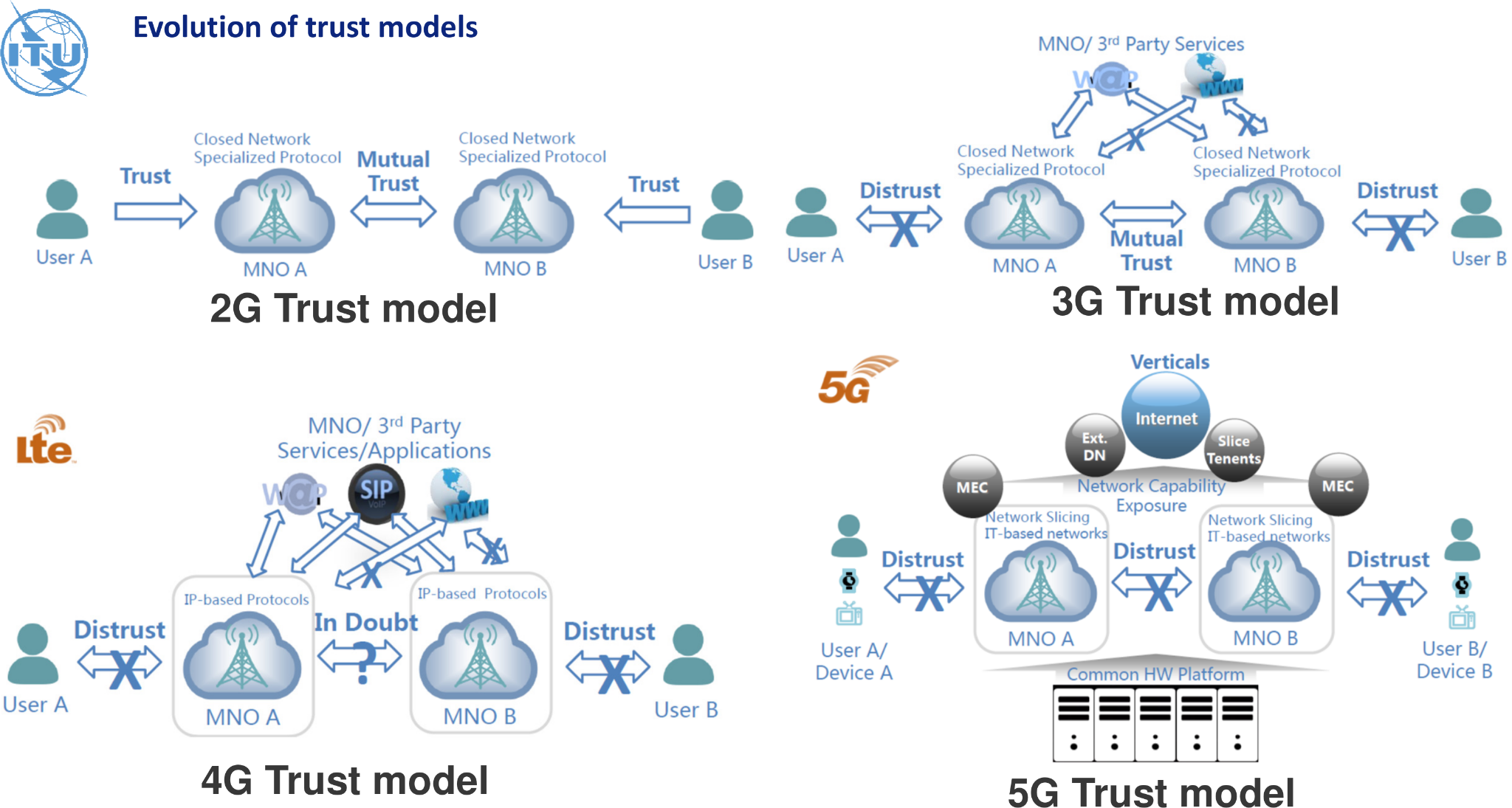}
    \caption{Evolution of trust models \cite{itu}.}
    \label{fig:trust}
\end{figure}

To address the continuous needs that have been pushed in the telecommunications ecosystem over the past three decades, the trust model of the various entities that are involved in the service provision has been radically redesigned. The continuous increase in the complexity has a big impact on the trust among the different nodes, which as expected, is continuously decreasing, see Figure \ref{fig:trust}. While it is not a zero-trust architecture, the amount of effort to decouple strong bonds among peers has reached a point where minimum trust is shared among different nodes and roles. In practice, mobile network operators are expected to split their networks into individual trust zones. Thus, subnetworks of other operators are assumed to belong to different trust zones, therefore, minimising possible leaks and exposure.

In the next paragraphs, we discuss the security of 5G networks from two perspectives. First, we present the building blocks that make 5G secure, which essentially provide the theoretical security of the network, that is, the encryption and integrity algorithms and protocols. Next, we discuss the privacy features of 5G. Then, we present the implementation aspect, which is focused on how vendors are actually controlled so that they deliver secure implementations of the network devices. Finally, we present an overview of the state of the art and practice attacks on 5G networks in the literature. 

\begin{figure}[th]
    \centering
    \includegraphics[width=\textwidth]{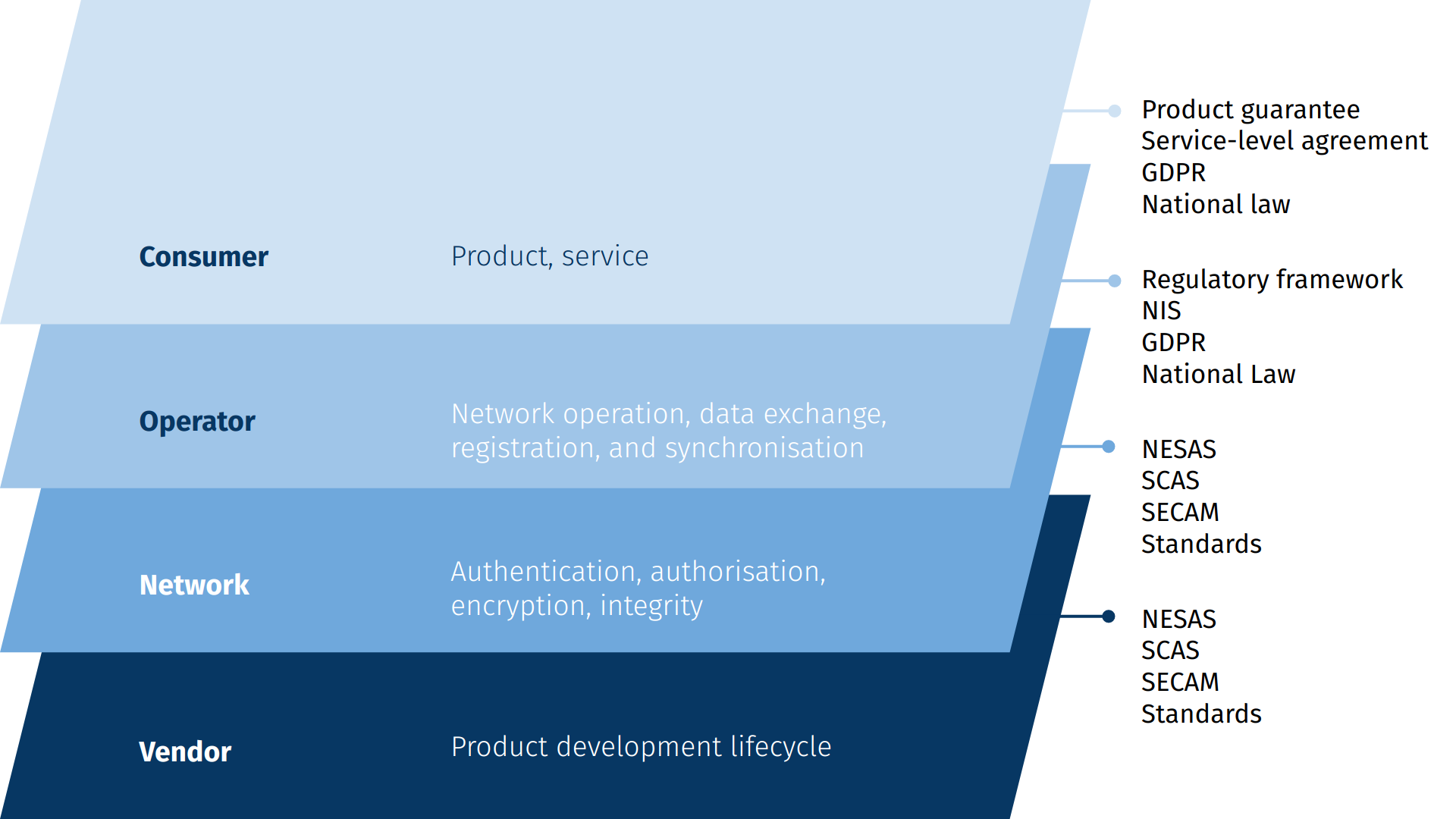}
    \caption{Entity-wise security guarantees}
    \label{fig:entity}
\end{figure}

In terms of guarantees, the operational part is not covered in this work, as the operator is monitored by the regulatory framework, e.g. NIS primarily and GDPR directives that were covered above. Finally, the user is covered by the GDPR, the service-level agreement (SLA), and national laws that protect customers. The above guarantees are illustrated in Figure \ref{fig:entity}.

\subsection{Network security}
\begin{figure}[th]
    \includegraphics[width=\textwidth]{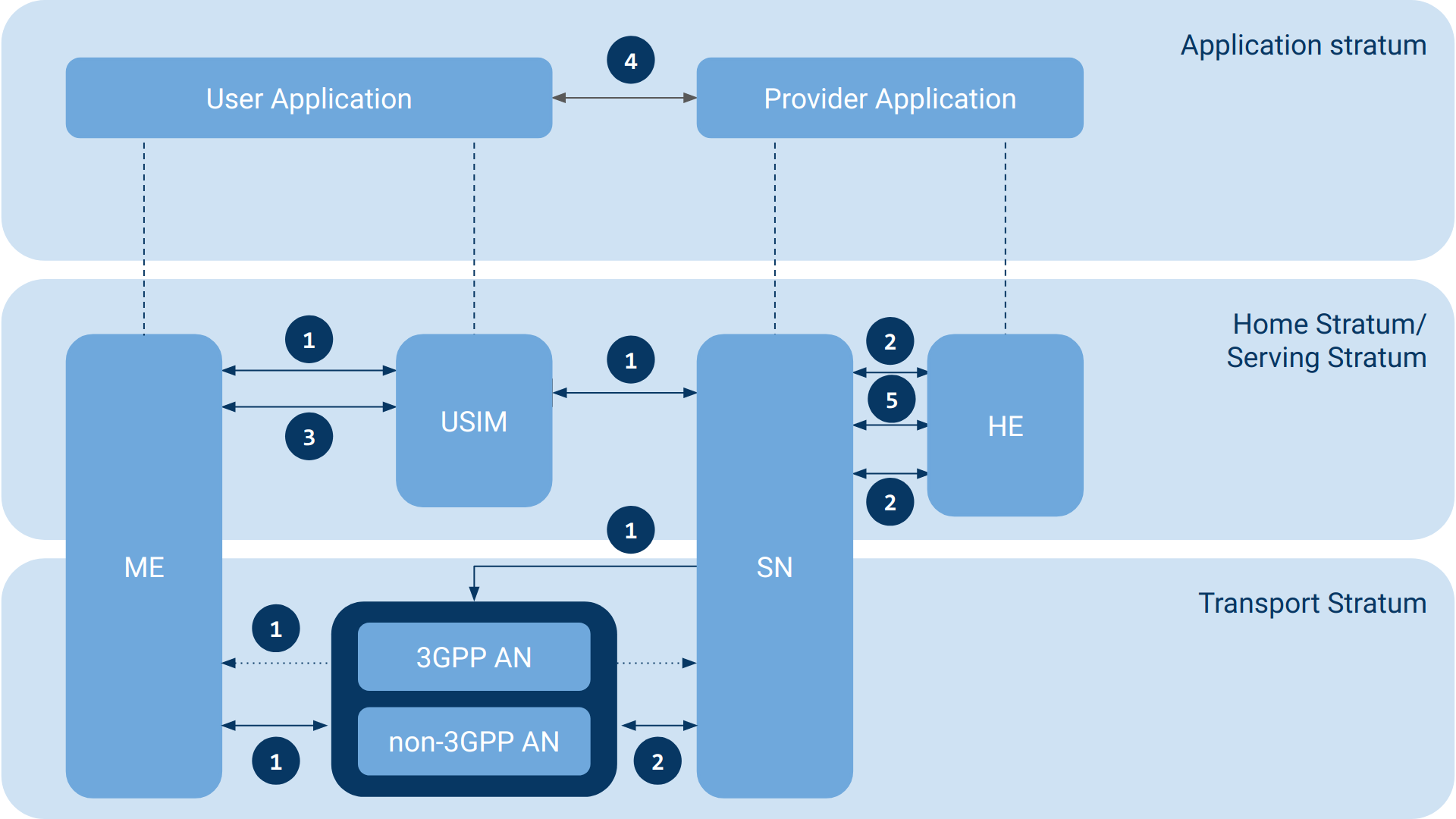}
    \caption{Security mechanisms of 5G in each layer.}
    \label{fig:stratum}
\end{figure}

To present the security architecture of 5G, we have to break down the stack into three layers, the application, home/serving, and transport layer, or stratum, as referred to in ETSI TS 133 501 \cite{3gpp2020security}. By doing so, we may categorise the security mechanisms into five distinct categories, which are numbered accordingly in Figure \ref{fig:stratum}:

\begin{itemize}
\item Network access security: Security features that enable a device to authenticate and access services via the network securely, mostly focusing on attacks on the radio interfaces and communicating the security context delivery from a Secondary Node to AN for access security.
\item Network domain security: the set of security features that enable network nodes to exchange signalling data securely, and user plane data.
\item User domain security: the security mechanisms that secure the user access to mobile equipment.
\item Application domain security: the security mechanisms that enable applications in the user and the provider domain to exchange information securely.
\item Service-Based Architecture (SBA) domain security: The security mechanisms that enable network functions of SBA to communicate within the serving network domain securely and other domains, including, but not limited to, registration, discovery, and authorisation, as well as the protection for the service-based interfaces. 
\item Visibility and configurability: the mechanisms that allow the user to determine whether a mechanism is working and to configure certain local security settings.
\end{itemize}

In terms of cryptographic primitives, in 5G, the choice is based on well-known, broadly used algorithms that have passed the strength of time criterion and are used by its predecessors. The symmetric encryption algorithms are called New radio Encryption Algorithm (NEA) and come in three flavours:
\begin{itemize}
\item 128-NEA1, a 128-bit SNOW 3G-based algorithm, identical to  128-EEA1;
\item 128-NEA2, a 128-bit AES-based algorithm \cite{dworkin2001advanced}  in CTR mode \cite{lipmaa2000ctr}, identical to 128-EEA2; and
\item 128-NEA3, a 128-bit ZUC-based algorithm, identical to 128-EEA3.
\end{itemize}
Note that the above 128-EEA* algorithms are defined in \cite{3gpp33401} and \cite{3gpp32221}, signifying the continuity with 5G's predecessors. 

Similar to NEAs, for integrity algorithms in 5G, we have the New radio Integrity Algorithms (NIA), which follow the same pattern, showing again the continuity with 5G's predecessors:
\begin{itemize}
\item 128-NIA1, a 128-bit SNOW 3G-based algorithm which is identical to 128-EIA1;
\item 128-NIA2, a 128-bit AES-based algorithm in CMAC mode, which is identical to 128-EIA2; and
\item 128-NIA3, a 128-bit ZUC-based algorithm which is identical to 128-EIA3.
\end{itemize}

Note that for both NEA and NIA there are "null" modes (named NEA0 and NIA0, respectively) which do not use  any cryptographic primitives which can be used in emergency calls and when establishing the security context of a typical attach procedure to access a cellular network  \cite{3gpp2020security}. This can be considered similar to the unencrypted cipher negotiation phase of TLS. However, in the early testing of 5G and misconfigurations have tricked the networks into using the null modes to launch attacks successfully \cite{rupprecht2016putting,rupprecht2019breaking,zhang2021towards}.
All computed hashes are based on SHA-256 \cite{national2002fips}, so for message authentication code, 5G uses HMAC–SHA-256. 

To generate secret keys from a provided value, the input to key derivation function (KDF) in 5G is defined in \cite{3gpp33220}. Nevertheless, the KDF is the IETF standardised HMAC-SHA-256 \cite{krawczyk1997ietf}.

For the authentication, 5G uses the well-known EAP-AKA' \cite{arkko2009rfc} (standardised by IETF as RFC 9048) and its enhanced 5G variant named 5G AKA, defined in \cite{3gpp2020security} as EAP-AKA' was also used in \cite{3gpp33401}. Finally, in private 5G networks, the default protocol is EAP-TLS, also standardised by IETF as RFC 5216.

For public key encryption, 5G uses the well-known Elliptic Curve Integrated Encryption Scheme (ECIES) \cite{abdalla1999dhaes}, which is a variation of the traditional ElGamal public key scheme. This scheme, for instance, is applied when User Equipment (UE) sends its SUPI over-the-air using the public key of the subscriber's home network which is stored the SIM card. The elliptic curve selection is rather optimised for speed as encryption and Diffie-Hellman key exchange use the safe elliptic curve Curve25519 \cite{bernstein2006curve25519,langley2016elliptic}, offering a bit less than the claimed 128 bits of security \cite{bernstein2016failures}.

All of the above have been used as the “binding glue” as building blocks to derive the proper protocols that can provide the necessary functionality securely. Moreover, these protocols have been formally analysed to determine their theoretical security \cite{basin2018formal,cremers2019component,zhang2020formal,peltonen2021comprehensive,edris2020formal} and specific provers have been developed \cite{hussain20195greasoner}. 

To further guarantee the security of the credentials and long-term keys by avoiding physical and software leaks, they are stored in tamper-resistant hardware in the UE. Moreover, the long-term keys must never be in any unencrypted form outside the UE's tamper-resistant hardware. Thus, any authentication algorithm that uses subscription credentials is executed by this hardware.

\subsection{User privacy}
In terms of privacy, it is clear that the target of attacks is the subscriber. To this end, the attacker would either try to disclose the identity of a user; which subsequently would lead to the disclosure of her location, or traffic monitoring to determine her actions on the network. To address these issues, 5G networks try to decouple the permanent identity of the user from the authentication phase and encrypt the traffic with secure cryptographic primitives. International Mobile Subscriber Identity (IMSI) catchers were a big threat for all 5G predecessors as the attacker could make the UE to transmit the identity of the user (IMSI) unencrypted, and subscribers in proximity would intercept it. 

\begin{figure}
    \centering
    \includegraphics[width=\textwidth]{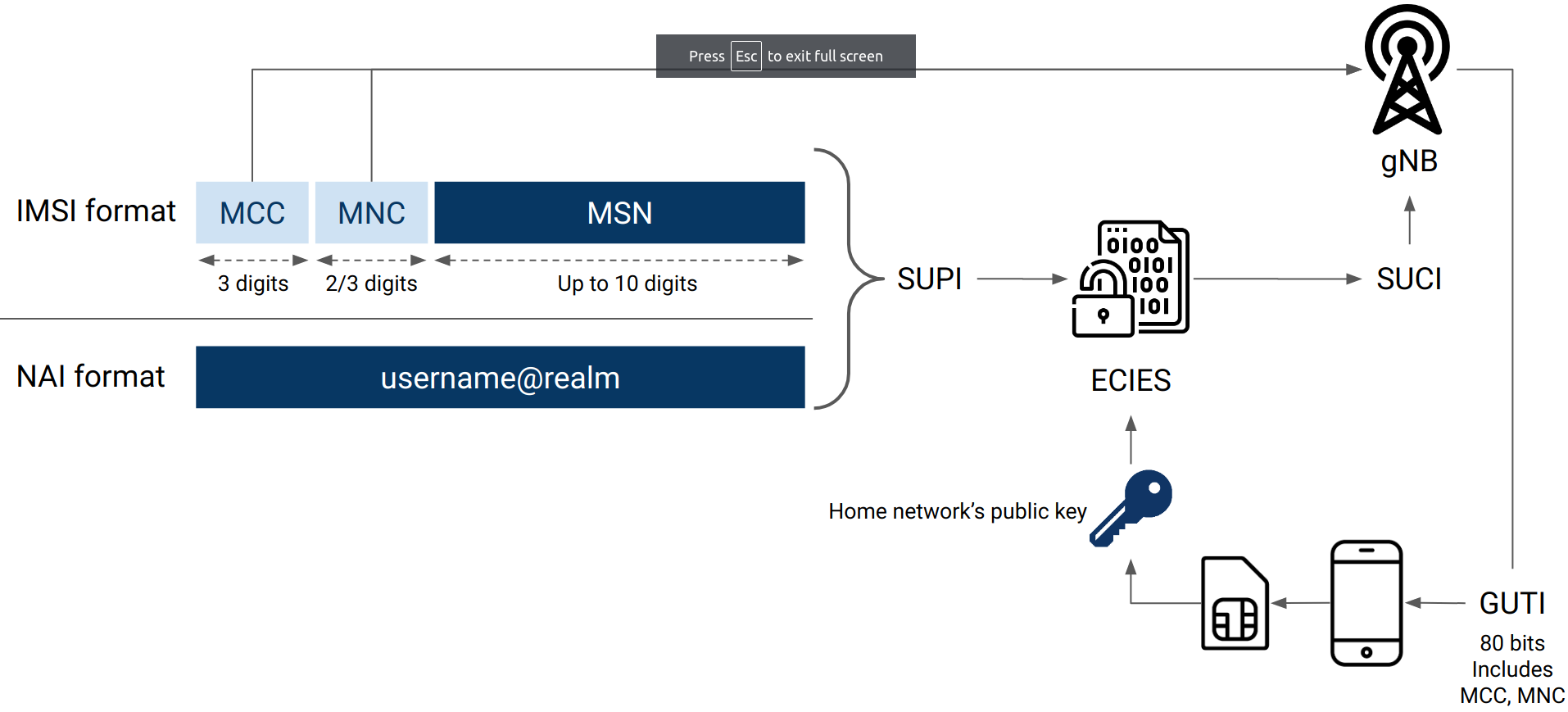}
    \caption{The SUPI format and encryption to SUCI.\\\textbf{Notation:} MCC: Mobile Country Code. MNC: Mobile Network Code. MSIN: Mobile Subscriber Identification Number.}
    \label{fig:supi_suci}
\end{figure}

In 5G, the equivalent of IMSI is the Subscription Permanent Identifier (SUPI) which is stored in each subscriber's USIM. It is either in IMSI format (up to 15 decimal digits) or as a
Network Specific Identifier (NSI) in the Network Access Identifier (NAI) format \cite{rfc4282}. Using ECIES and the public key of the Home Network in the USIM, UE encrypts SUPI (in the case of IMSI, a part of it) and creates the Subscription Concealed Identifier (SUCI). This way, each time UE has to authenticate, SUPI is never disclosed in cleartext. Finally, the visiting network may assign a new temporary identity to the UE called Globally Unique Temporary Identity (GUTI). This identifier is used to prevent eavesdroppers on the radio link from identifying UEs. Practically, GUTI is a temporary identifier that has to be frequently updated and used for identification over the radio access link. The process is illustrated in Figure \ref{fig:supi_suci}.

In the IMSI format, SUPI has a constant length and the same applies for SUCI. However, if the NAI format is used, the length is variable and deanonymisation attacks can be performed \cite{preuss2021nori}. Notably, while IMSI-catcher attacks are countered by 5G-AKA, other privacy 5G-AKA attacks remain \cite{8806761}. Moreover, while GUTIs are a very good privacy measure, their update is subject to the network configuration. Therefore, if the network is not properly configured, GUTIs can remain unchanged arbitrarily long and ultimately serve as accurate user identifiers.

For more on privacy aspects of 5G, the interested reader may refer to \cite{khan2018defeating} and \cite{khan2020survey}.

\subsection{Vendor-wise security}
A set of frameworks have been established to assess the security of 5G networks in an objective and replicable way. The most broadly accepted and used framework is by 3GPP and GSMA. In essence, due to the highly heterogeneous environment that 5G has, there is an attempt to break down the proposed security architecture by referring to other standards. The core idea is that by splitting the architecture and isolating layers, modules, and components, one may objectively assess the security of each mechanism individually. While the fact that each part of the whole is individually secure does not guarantee the security of the whole, the task is significantly reduced as one has to check the interfaces that glue this together. Therefore, instead of having a holistic approach that might be 5G specific but needs to start from scratch, 3GPP and GSMA favoured the onion-layer approach by using the broadly accepted security mechanisms from ETSI, IETF, and ITU-T. This way, the corresponding stakeholders may refer directly to the corresponding mechanism and roll out the measures directly. For instance, Network Function Virtualization and edge security were based on the work from ETSI, while SDN and cloud computing security were based on ITU-T. Based on the above, the two major questions that arise are the following:

\begin{itemize}
\item How do we certify that the necessary security mechanisms have been properly implemented? After all, the specifications of a mechanism might be correct; however, the implementation might be flawed.
\item How do we certify that the operation of the 5G network is secure and that the security mechanisms that have been commissioned are actually operated by the corresponding service provider? In this case, the question is whether the operator makes use of the security mechanisms 
\end{itemize}

Because of the nature of these questions, the answer is not unique and there cannot be a sole mechanism to monitor both as they refer to radically different stakeholders operating in different jurisdictions, so different checkpoints must be established. In this regard, it is better to model these interactions as part of a supply chain. To this end, we have the vendors of the 5G equipment that push the products in the market to consumers, practically end-point devices, and operators where the products are part of the 5G backbone infrastructure. 

The case of consumers is rather straightforward as the consumer has the traditional means to test and verify the quality of the product she purchased. For the case of vendors and operators, due to the heterogeneous nature of the 5G ecosystem and its criticality, the checkpoints have to be split. In what follows, we first investigate the case of vendors.

Unfortunately, common practice has shown that the integration of security mechanisms in products has been done wrong far too many times and that products reach the market with several inherent vulnerabilities, poorly implemented methods or even a lack of provision for such mechanisms. Common issues such as default credentials or embedded on the firmware have been plaguing internet-facing devices for years. After the introduction of Zmap \cite{durumeric2013zmap}, which enabled internet-wide scanning, Shodan \cite{bodenheim2014evaluation} identified numerous internet-faced devices that should not be connected to the internet, many of which used default credentials. Mirai \cite{antonakakis2017understanding}, for instance, took advantage of the latter and exploited default router credentials to create a massive botnet that brought many high-profile services to their knees with its denial of service \cite{miraicloud,guardian_2016}. Given the continuous revelation of such security issues and that often applying security patches, even if a device is not compromised, significantly impacts the quality of service or may not always be possible, the problem is augmented in the 5G context. Especially for the infrastructure, trust in a device's security becomes a paramount priority. 

To address this issue, 3GPP and GSMA have opted to follow a continuous iteration approach, which, even if not directly stated, follows the DevSecOps approach. DevOps is a set of practices that combine the application of Agile and Lean approaches to operations work. It also includes the collaboration between development and operations staff throughout all stages of the software development lifecycle (SDLC) and information-technology operations. DevOps aims to shorten the SDLC and provide continuous delivery with improved software quality. Serverless computing, API-first applications, and microservices-based architectures are a baseline in software development and deployment, and DevOps is the main enabler for these paradigms. Unfortunately, these new approaches have evolved without security being a well-thought-out design principle from the very beginning. The only real option available to inject security in a sustainable manner into these recent and complex paradigms is to extend their very core and include it from the start. This natural extension, DevSecOps, reduces the complexity of software development and deployment, ensuring that only known and trusted components and services are used. DevSecOps enables features such as empowering development teams with security resources, integrated directly into their development methods, using well-secured and monitored environments and ultimately delivering a product with improved overall quality. 

In this train of thought, GSMA has collaborated with 3GPP, and their DevSecOps approach is actualised through a two-face approach that complements each other; one faces products while the other faces vendors. From the 3GPP side, we have the Security Assurance Methodology (SECAM), and from the GSMA side, the Network Equipment Security Assurance Scheme, commonly referred to as NESAS. SECAM sets the product security requirements and test specifications that will be performed on the products, commonly referred to as Security Assurance Specifications (SCAS), and how the evaluation will be performed. NESAS, on the other hand, defines the vendor processes requirements and audit methodology, defines who can test the products, and intervenes to resolve disputes. As a result, we have a supply chain security framework which provides high guarantees that the 5G products delivered by network equipment vendors will conform to high-security standards. Thus, further than merely receiving a certification that a specific product has passed some security checks, this approach investigates the whole lifecycle of the product. The whole lifecycle is illustrated in the figure below. 

3GPP defines the features and capabilities of a class of network products. Once they are defined, a SCAS first describes the security requirements for such products as different classes are exposed to different attacks and different threat models have to be considered. Then, a list of the tests that must be performed is provided. Clearly, a SCAS is not tied to a specific vendor and product but refers to a class of products that do not impede fair competition or impose further constraints on vendors. At the time of writing, the coverage of SCAS for 16 classes of products has been published, e.g. Mobility Management Entity (MME - TS 33.116) and eNodeB (eNB - TS 33.216), and the catalogue of general security assurance requirements (TS 33.117).

The testing of products based on the corresponding SCAS is performed by a set of security test laboratories which have been certified with ISO/IEC 17025. The latter guarantees that the baseline of security checks is of very high quality. While the process to join is open, currently, there are only seven authorised GSMA NESAS test laboratories. Moreover, there are two GSMA NESAS Appointed Auditors to carry out the Process Audit. GSMA publicly shares all process audits and product evaluations for transparency. As a result, this approach has managed to receive a quick market uptake and reached a consensus with vendors, as the vast majority of them are also members or participate in 3GPP and GSMA. 

\begin{figure}[th]
    \centering
    \includegraphics[width=\textwidth]{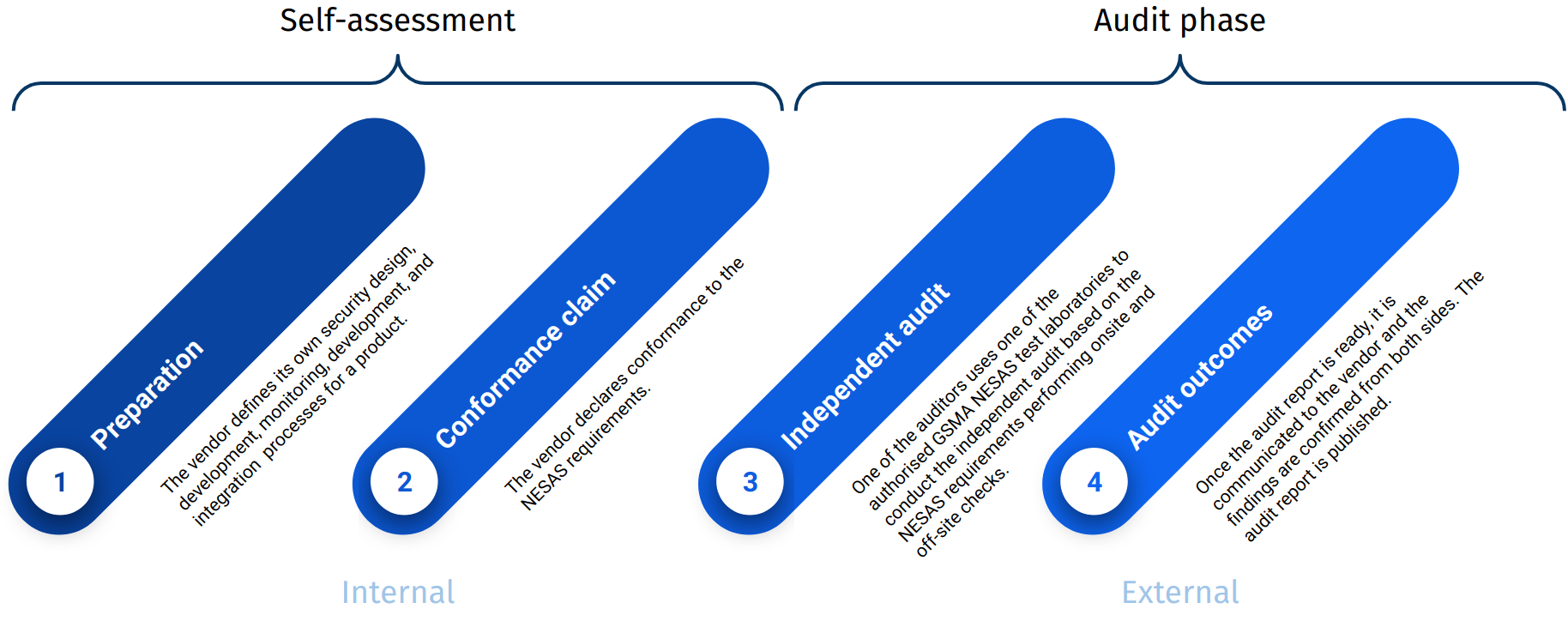}
    \caption{The NESAS lifecycle.}
    \label{fig:nesas}
\end{figure}

The process for monitoring the development lifecycle of a vendor is illustrated in Figure \ref{fig:nesas}. Practically, each vendor has its own processes for developing a product and bringing it to the market, which may vary greatly. Therefore, complete documentation and check of the internal processes are conducted by each vendor. However, this is a common practice and requirement for many, e.g. ISO certifications, so the overhead is not significant and does not clash with other processes. Once a vendor assesses that it is mature enough to meet the NESAS requirements, it files an audit request. While this part was solely internal, the audit request now positions the vendor to be independent and external security checks. Therefore, GSMA appoints one of the Auditing Organisations to conduct the audit. The audit is conducted by one of the authorised GSMA NESAS test laboratories that perform the NESAS methodology to determine the vendor's conformance to the corresponding requirements. Once the checks are performed, the report is assembled and communicated to the vendor to validate the results. The consensus report is then published and open to everyone. The independence of the organisations and their conflicting interests guarantees that the positive outcome of this process implies that the vendor indeed follows high standards in its product development lifecycle.

\begin{figure}[th]
    \centering
    \includegraphics[width=\textwidth]{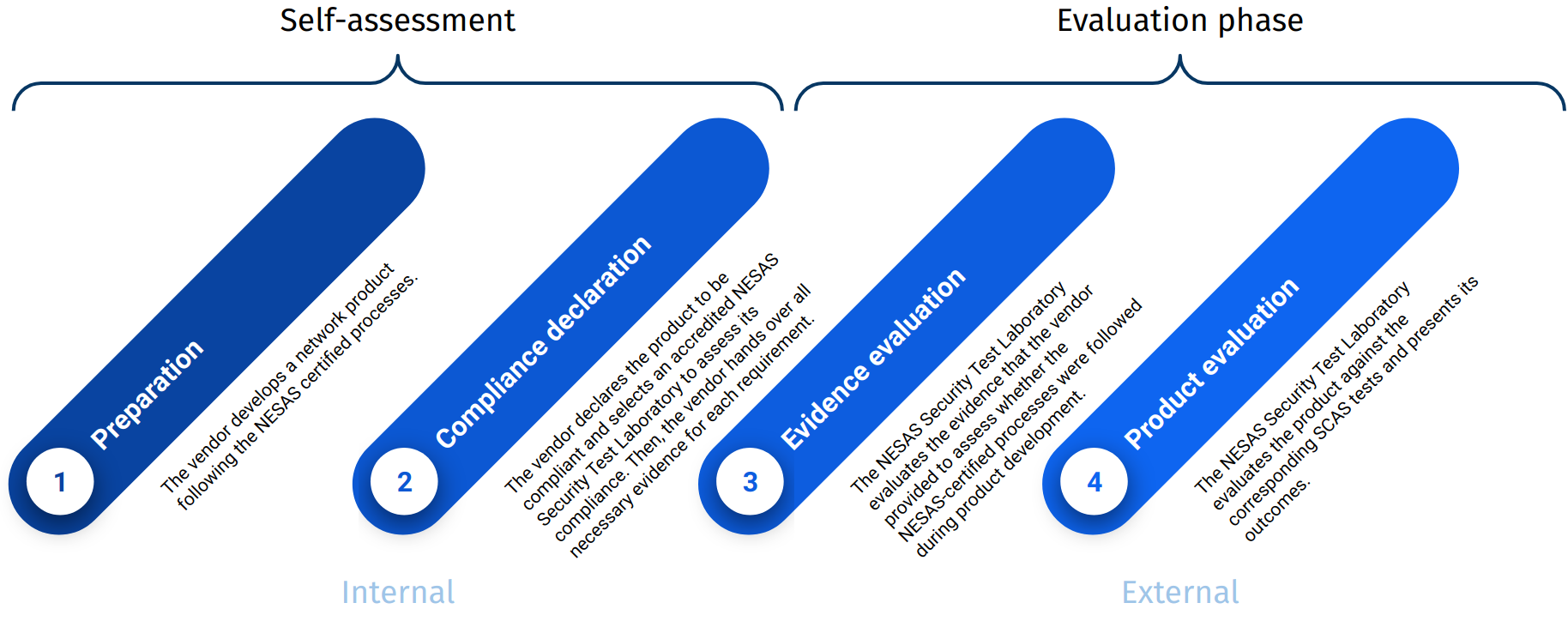}
    \caption{Testing a product with SCAS.}
    \label{fig:scas}
\end{figure}

Once a vendor is verified to conform to the NESAS requirements, it may proceed to verify network products. This guarantees that the products that would be verified to conform to the security requirements have been produced by vendors that follow very strict and security-aware processes. As a result, the guarantees are much stronger than having a product that complies with, e.g. Common Criteria. Essentially, the process for the products is very similar to the one for NESAS compliance, see Figure \ref{fig:scas}. Practically, we have two phases, self-assessment and evaluation. In the first phase, the vendor develops a network product according to the NESAS-certified processes. Once the vendor considers that the product conforms to the security requirements of the corresponding product class, it chooses one of the NESAS Security Test Laboratories to conduct the independent audit on the product and provides all the necessary compliance evidence for the product, as internally assessed. Then, the Test Laboratory evaluates the evidence that the vendor provided to assess whether the NESAS-certified processes were followed when developing the evaluated network product. Having determined that the process was followed, the corresponding SCAS tests are performed on the device and a report is delivered to present its outcomes. Again, the competing interests of the vendor and the test lab can guarantee that the product that passes this process will conform to high-security standards and will not be vulnerable to known attacks.

While many of the above may sound very bureaucratic and lack technical details, it is crucial to understand whether the process works. In short, there is no disclosed vulnerability affecting certified devices. Based on the fact that 5G is in the spotlight and that hackers who could lay their hands on such equipment or market competitors would immediately disclose such findings, it is fair to say that the process works well. Of course, there is no such thing as absolute security; nevertheless, from the so far experience, one can argue that the bar is rather high. Yes, there are some 5G-relevant attacks, see next section; however, their impact on the core infrastructure as they mostly stem from configuration issues that can be resolved. Therefore, their actual impact is not significant enough to pose an actual threat at the moment. 

Finally, it is worth reporting that the GSMA-3GPP approach is not the only one. For instance, the Computer Emergency Response Team (CERT) of the Organisation of The Islamic Cooperation (OIC) \cite{oic} has recently initiated a framework called OIC 5G Security Framework based on open standards. Due to the amount and size of stakeholders backing this framework, it is interesting to see how this initiative will evolve. While the approach is interesting, it has not reached the maturity of the GSMA-3GPP approach, which is also based on open standards \cite{cheang2021achieving}. Nevertheless, the authors argue that convergence and merge of such approaches might be the best option. This would avoid having splits that could lead to products having to conform against possibly contradicting rules or having two different markets for the same infrastructure. Common practice has shown that living “between two kings”, or even worse, multiple, impedes honesty to at least one of them.

\subsection{Real-world attacks}
As in every domain, there is no such thing as absolute security. Over the past few years that 5G has been rolled out, even experimentally, security researchers have put 5G in the spotlight and developed several attacks. However, some of them are due to issues with the specifications. Compared to the size of 5G infrastructure, its massive infrastructure and code base, while important and relevant, they cannot be considered a serious threat to the ecosystem. Moreover, the bulk of the attacks are against end users' equipment service provision and privacy, so they do not reach the core systems.

We categorise the attacks as follows:
\begin{itemize}
    \item IMSI catching is a man-in-the-middle attack where the attacker uses a fake antenna to intercept user traffic.
    \item Device, web, and app fingerprinting: These attacks aim to identify devices and applications used by the victim or traffic to specific websites. 
    \item Location tracking: In these attacks, the attacker tries to extract the victim's location by collecting the SUPI. In this context, one cannot extract the exact location but the proximity to an antenna.
    \item Downgrade: The attacker aims to downgrade the victim's connection from 5G to, e.g. 4G, decreasing the service quality.
    \item Denial of Service: In these attacks, the attacker tries to prevent access to a specific service by blocking messages, draining resources, etc.
    \item Eavesdropping: Attackers try to obtain access to data exchanged between the victim and an unprotected network.
    \item Impersonation: The attacker impersonates the victim in the network to perform actions in its name.
    \item Data manipulation: The attacker manipulates the data that the victim exchanges, and the victim cannot identify the alterations.
    \item Warning: The attacker issues fake warning notifications to the subscribed users to spread fake news of an upcoming threat. 
    \item Protocol/specifications: In this category, we have attacks that have discovered vulnerabilities in the protocols and specifications of 5G.
\end{itemize}
Table \ref{tbl:attacks} presents the relevant attacks to real-world 5G networks. To address these issues, a continuous stream of research is devoted \cite{singla2021look,ren2021fast,singla2020protecting,ettiane2021toward,behrad2020new,thantharate2020secure5g,apruzzese2022wild,sec5g}, but also updates to configurations and implementations to protect the end-users.

For more details, the interested reader may refer to \cite{9491923,jover2019security,shaik2019new,zhao20215g,yu2021improving,park20215g,tang2022systematic,ramezanpour2022security,bang2022opportunistic,khan2019survey}.

\begin{table}[th]
\centering
\begin{tabular}{lll}
\toprule
\textbf{Attack} &\textbf{Type} & \textbf{References} \\ \midrule
Alerts & Security & \cite{bitsikas2022you}\\
Device, web, and app fingerprinting & Privacy & \cite{rupprecht2016putting,rupprecht2019breaking,shaik2019new,zhai2021identify} \\
Denial of Service & Security & \cite{shaik2019new,hussain20195greasoner,bitsikas2022you,ludant2021sigunder,yang2019hiding,adaptover,bitsikas2021don} \\
Downgrade & Security/Quality& \cite{hussain20195greasoner,shaik2019new} \\
Eavesdropping & Security/Privacy & \cite{rupprecht2016putting,fang2017security,zhang2021towards,xu2021qos,xu2020incentive,sanchez2021survey,adaptover,RupprechtKHP20,bitsikas2021don} \\ 
Data manipulation & Security & \cite{rupprecht2019breaking,ludant2021sigunder,adaptover}\\
IMSI catching & Privacy & \cite{rupprecht2019breaking,hussain2019privacy,rupprecht2016putting,chlosta20215g}  \\
Impersonation & Security & \cite{chlosta2019lte}\\
Location tracking & Privacy& \cite{rupprecht2016putting,zhang2021towards,shaik2019new,lakshmanan2021stealthy} \\
Protocol/specifications & Security/Privacy & \cite{hussain20195greasoner,8817957,bitsikas2021don}\\
\bottomrule
\end{tabular}
\caption{Real world attacks to 5G networks.}
\label{tbl:attacks}
\end{table}

Note that some of the attacks, e.g. \cite{rupprecht2016putting,rupprecht2019breaking,adaptover}, constitute multiple attacks according to our categorisation.

\section{Conclusions and the road ahead}
5G marks a new era for telecommunications and computing as a whole. As discussed, it is not that 5G subscribers will enjoy faster connection speeds, lower latency, better reliability, and increased availability on their devices. The main changes come from the plethora of applications that 5G seamlessly enables and can radically improve the quality of services, products and, as a result, the quality of everyday life. From its inception, 5G fostered a security approach that manages to fix many issues of its predecessors. As a result, the inherent security measures provide high-security guarantees from which users can benefit. Going a step beyond, one may question whether these security guarantees are claimed or actually implemented and how this can be verified. To this end, in this work, we provided an overview of how manufacturers prove their conformance to these standards in a transparent, highly monitored, and auditable form that many manufacturers have fostered, using the NESAS and SCAS way of GSMA and 3GPP. The market endorsement of this model as well as the conformance to existing and open standards can only benefit the whole ecosystem. 

The recent publication of the Commission's adoption of the new Cyber resilience act \cite{crct} seems to be well-aligned with the processes established by GSMA and 3GPP, so the compliance of, e.g. NESAS v3.0 will not require major updates. Nevertheless, due to the continuous integration of AI and ML in current technologies, we argue that the upcoming AI Act \cite{aiact} should take into consideration 5G as beyond an enabler for many AI and ML applications, 5G also makes use of these technologies. The above illustrates that despite NESAS and SCAS not having many years on their back, it has matured enough to be one of the cornerstones of seamlessly implementing 5G security measures.

We argue that due to the continuous developments in quantum computing and the recent adoption of quantum-resistant cryptographic algorithms from NIST\footnote{\url{https://www.nist.gov/news-events/news/2022/07/nist-announces-first-four-quantum-resistant-cryptographic-algorithms}}, steps for the post-quantum era should be made. The impact that quantum computing could have on 5G is detrimental \cite{mitchell2020impact} as for instance, elliptic curve primitives are not quantum-resistant, and the symmetric key primitives will lose at least half of their key-length security. Table \ref{tab:alg_security} shows the current and post-quantum security of the cryptographic primitives that 5G uses. Notably, only the underlying hashing algorithm can be considered quantum resistant. Currently, most initiatives to address this gap are from the academia \cite{chamola2021information,ulitzsch2022post,clancy2019post,10.1145/3507657.3529657} yet this discussion must soon reach the rest of the stakeholders to mitigate possible future risks.

\begin{table}[th]
\centering
\begin{tabular}{@{}lrr@{}}
\toprule
\textbf{Primitive} & \textbf{Bits of security} & \textbf{PQ bits of security} \\ \midrule
AES-128 & 128 & 64 \\
SNOW-128 & 128 & 64 \\
ZUC-128 & 128 & 64 \\
SHA-256 & 256 & 128 \\
Curve25519 & 126 & 0 \\ \bottomrule
\end{tabular}
\caption{The current security level of 5G cryptographic primitives and in the post-quantum era.}
\label{tab:alg_security}
\end{table}

Finally, while we acknowledge that backward compatibility may provide many functional guarantees, many of the 5G attacks stem from attacks on its predecessors. Therefore, it is essential to revise which functionalities of its predecessors must be ported and in which context to minimise the threat surface. Similarly, the use of null algorithms should also be reconsidered, as they are often abused to launch attacks that trick UEs and networks into using them.

\section*{Acknowledgements}
This work was supported by the European Commission under the Horizon Europe Programme, as part of the project \textit{LAZARUS} (Grant Agreement no. 101070303).

The content of this article does not reflect the official opinion of the European Union. Responsibility for the information and views expressed therein lies entirely with the authors.

\bibliographystyle{plain}
\bibliography{refs}

\end{document}